\documentclass[conference, a4paper]{IEEEtran}
\usepackage[a4paper, top=1.9cm, bottom=4.4cm, left=1.6cm, right=1.6cm]{geometry}
\IEEEoverridecommandlockouts
\usepackage{cite}
\usepackage{amsmath,amssymb,amsfonts}
\usepackage{algorithmic}
\usepackage{graphicx}
\usepackage{textcomp}
\usepackage{xcolor}
\usepackage{multirow}
\usepackage{enumerate}
\usepackage{bbding}
\usepackage{enumitem}
\def\BibTeX{{\rm B\kern-.05em{\sc i\kern-.025em b}\kern-.08em
    T\kern-.1667em\lower.7ex\hbox{E}\kern-.125emX}}

\begin{document}

\title{Near-Linear Scaling Data Parallel Training with Overlapping-Aware Gradient Compression\\}

\author{
    \IEEEauthorblockN{Lin Meng$^{*\dag}$, Yuzhong Sun$^{*}$, Weimin Li$^{*\dag}$}
    \IEEEauthorblockA{$^*$ Institute of Computing Technology, Chinese Academy of Sciences}
    \IEEEauthorblockA{$^\dag$ University of Chinese Academy of Sciences}
    \IEEEauthorblockA{\{menglin20z, yuzhongsun, liweimin19z\}@ict.ac.cn}
}

\maketitle

\begin{abstract}
Existing Data Parallel (DP) trainings for deep neural networks (DNNs) often experience limited scalability in speedup due to substantial communication overheads. While Overlapping technique can mitigate such problem by paralleling communication and computation in DP, its effectiveness is constrained by the high communication-to-computation ratios ($CCR$) of DP training tasks. Gradient compression (GC) is a promising technique to obtain lower $CCR$ by reducing communication volume directly. However, it is challenging to obtain real performance improvement by applying GC into Overlapping because of (1) severe performance penalties in traditional GCs caused by high compression overhead and (2) decline of Overlapping benefit owing to the possible data dependency in GC schemes. In this paper, we propose COVAP, a novel GC scheme designing a new coarse-grained filter, makes the compression overhead close to zero. COVAP ensures an almost complete overlap of communication and computation by employing adaptive compression ratios and tensor sharding tailored to specific training tasks. COVAP also adopts an improved error feedback mechanism to maintain training accuracy. Experiments are conducted on Alibaba Cloud ECS instances with different DNNs of real-world applications. The results illustrate that COVAP outperforms existent GC schemes in time-to-solution by 1.92x-15.39x and exhibits near-linear scaling. Furthermore, COVAP achieves best scalability under experiments on four different cluster sizes.
\end{abstract}

\begin{IEEEkeywords}
distributed deep learning, Overlapping, gradient compression, data parallel
\end{IEEEkeywords}

\section{Introduction}
With an increasingly large cluster of GPU nodes, Data Parallel training (DP) has become one of the most widely adopted norms to support large-scale Deep Neural Network (DNN) training \cite{b12}. In typical settings of DP, each worker uses the same model to do the forward and backward pass on its own data partition. After the backward pass (BP), the local gradients of each worker are exchanged in the process group \cite{b8}.

In the optimal case of DP, each worker iterates over its own data partition at the same speed as local training. Since the large dataset is divided into N parts (N represents the number of workers), the overall training time is also reduced by N times. This optimal case is called linear scaling \cite{b4,b8}. However, due to the non-negligible communication overhead caused by gradient synchronization in typical computing environments (e.g., cloud computing), the speedup of default DP is far from linear scaling. Fig. 1(a) shows different phases in one iteration of DP. The communication is relatively long due to the communication bottleneck of DP \cite{b25}.

Several system optimizations \cite{b14, b21} try to utilize the opportunity of paralleling computation and communication to alleviate such communication bottlenecks. We call these methods uniformly as Overlapping. Overlapping is based on the independence of DNN’s gradient calculation, which means gradients are calculated layer by layer. Therefore, instead of waiting to complete all gradients’ calculations, current deep learning frameworks \cite{b1,b18} start communications in advance once some of the gradients is ready. Then, the computation of the following parts of gradients can be executed in parallel with the communication of the above layers’ gradients. Fig. 1(b) shows how Overlapping is exploited in one iteration of DP training.

In optimistic cases, the computation time is longer or similar to the communication overhead, so Overlapping can hide most of the communication overhead and makes DP close to linear scaling. However, the computation time and communication overhead of different DNNs varies significantly due to the massive difference in the number of parameters and model structures. Table \uppercase\expandafter{\romannumeral1} shows the computation time and communication overhead of several commonly used DNNs in our environment. Since the communication overhead is usually much larger than the computation time, the speedup provided by Overlapping (denoted by $S_{ovlp}$) is considerably lower than the ideal speedup of linear scaling (denoted by $S_{LS}$). $T_{before}$ represents the computation time before the backward pass, including forward pass and data loading.
\begin{figure}[t]
\setlength{\abovecaptionskip}{0.cm}
\setlength{\belowcaptionskip}{-0.cm}
  \centering
  \includegraphics[width=86.5mm]{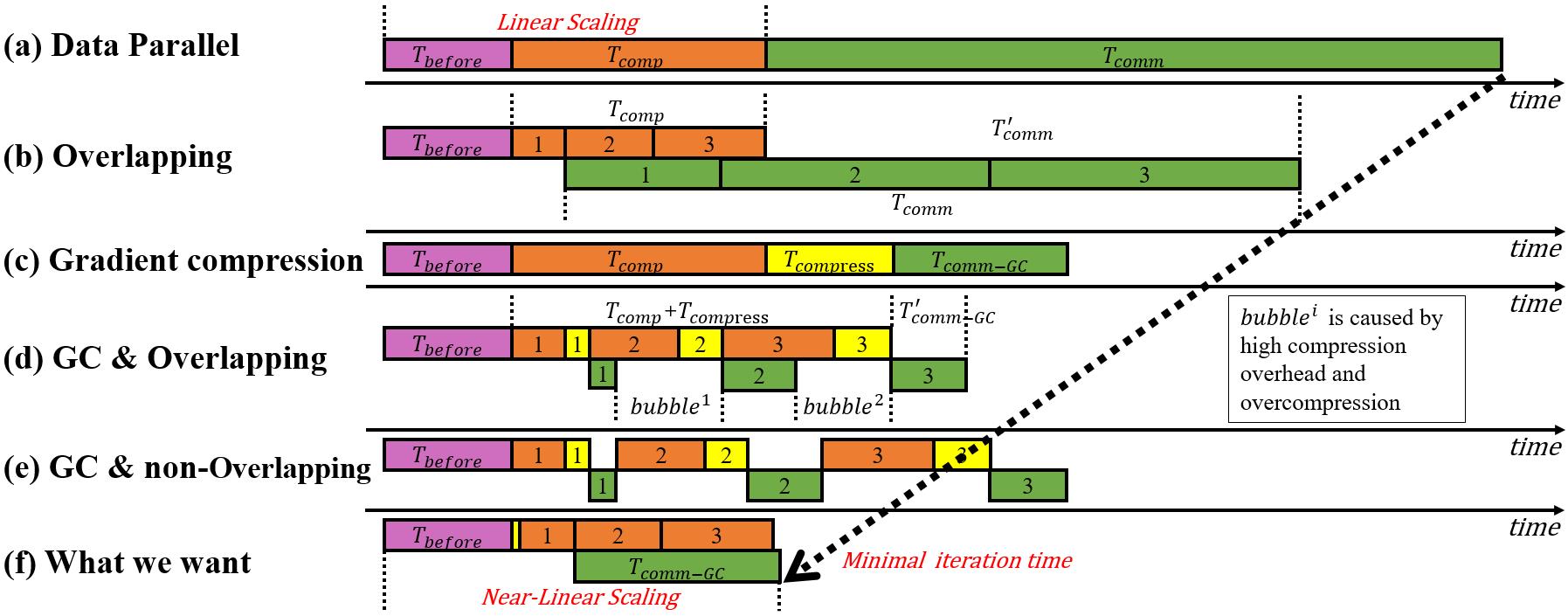}
  \caption{Examples of phases in DP with different train strategies. Linear scaling in (a) represents the optimal case. (f) represents an ideal case with almost fully overlapping and near-zero compression overhead. The decompression operations are omitted here.}
\end{figure}

Regardless of Overlapping, another more direct way to alleviate the communication bottleneck in DP is gradient compression (GC). GC generally falls into two categories: sparsification and quantization \cite{b5}. Sparsification utilizes the sparsity of gradients and only transmits a small part of gradients by some kind of filter (e.g., Top-k \cite{b3} selects the largest k gradients), while quantization decreases the precision of all gradients (e.g., transform the data type from FP32 to FP16). Both ways can significantly reduce the data volume of communication. However, the speedups provided by GC schemes in practice are far behind expectations \cite{b2,b4,b28}. One reason is the non-negligible compression overhead. Besides, GC operations generally start between gradients’ computation and communication. Such time dependency makes the overhead of GC difficult to be amortized \cite{b4}. Fig. 1(c) shows the timeline of GC.

\begin{table}[t]
\centering
\setlength{\abovecaptionskip}{0.cm}
\setlength{\belowcaptionskip}{-0.cm}
\setlength{\leftskip}{-2pt}
  \caption{The computation times and communication overheads of different DNNs}
  \resizebox{86.5mm}{7mm}{
  \begin{tabular}{ccccccc}
    \hline
    DNN&$T_{before}$&$T_{comp}$&$T_{comm}$&$CCR$&$S_{ovlp}$&$S_{LS}$\\
    \hline
    ResNet-101&55ms&135ms&280ms&2.1&1.43x&2.47x\\
    VGG-19&105ms&210ms&842ms&4.0&1.22x&3.04x\\
    Bert&80ms&170ms&520ms&3.1&1.28x&3.08x\\
  \hline
  \end{tabular}}
\end{table}

Considering GC and Overlapping simultaneously, our key insight is that using GC to obtain lower communication overhead while using Overlapping to parallelize communication and computation can reduce most communication overhead, as shown in Fig. 1(d). However, a potential challenge to combining these two methods is that some GC schemes contain extra synchronized communication operations, and subsequent computing operations (i.e., gradient computation and compression) rely on the communication results, which force the following computing to wait for completion of communication. This problem is called data dependency in GC. Fig. 1(e) shows an example of data dependency among computations, compressions and communications when using GC and Overlapping concurrently. The root cause is that existent GC schemes neglect the combination of Overlapping in their design.

Furthermore, to fully unleash the advantages of GC, the next challenge is to reduce compression overhead as much as possible. Existent GC schemes may contain some operators with high time complexity. For example, most of the overhead in Top-k \cite{b3} comes from {\itshape topk()} operator. Experiments are conducted to measure the compression overheads of existent GC schemes in Table \uppercase\expandafter{\romannumeral2}. Those overheads vary hugely from 5ms to 1560ms, which severely impairs the advantage of GCs. We observe that lower compression overhead generally leads to faster training. In optimal cases, the compression overhead should be near zero.

In this paper, we propose a new GC scheme called COVAP to address those challenges in order to make DP achieve near-linear scaling by using GC and Overlapping concurrently. We make the following contributions:
\begin{itemize}[itemsep=0pt,topsep=0pt,parsep=0pt,leftmargin=10pt]
    \item We propose a novel Overlapping-aware GC scheme: COVAP, which uses a new coarse-grained filter to compress gradients with near-zero overhead. We further design a tensor sharding technique to balance tensor sizes and avoid communication bottlenecks on large tensors. The error feedback \cite{b24} is also employed to guarantee convergence and preserve training accuracy.
    \item We design a new strategy to select compression ratio according to communication-to-computation ratio ($CCR$) in different training tasks, which consistently ensures almost complete overlap between computation and communication. Contrary to the large and constant compression ratio in traditional GCs, our design adopts much smaller ratios and automatically adjusts them according to variant DNNs and computing environments.
    \item We implement COVAP in PyTorch distributed data parallel module (DDP) \cite{b14}, which can be integrated using the communication hook. Our code is available at https://github.com/Sun-Helloworld/COVAP. Experiments show that COVAP improves training speed significantly by up to 15.39x compared to popular GC schemes while achieving almost no loss of accuracy. For scalability, COVAP achieves near-linear scaling on four clusters of 8, 16, 32 and 64 GPUs.
\end{itemize}

\begin{table}[t]
\setlength{\abovecaptionskip}{0.cm}
\setlength{\belowcaptionskip}{-0.cm}
\setlength{\leftskip}{-2pt}
  \caption{The compression overheads and communication time reductions of different GC schemes in VGG-19 training}
  \label{tab:freq}
  \resizebox{86.5mm}{12.5mm}{
  \begin{tabular}{cccc}
    \hline
    GC schemes&Hyperparameter&$T_{compress}$&$T_{comm}-T_{comm-GC}$\\
    \hline
    Top-k \cite{b3}&k=1\% &1560ms&603ms\\
    DGC \cite{b16}&k=0.1\%&25ms&747ms\\
    Random-k \cite{b23}&k=1\%&200ms&653ms\\
    FP16&---&~5ms&423ms\\
    EFsignSGD \cite{b12}&---&20ms&-210ms\\
    PowerSGD \cite{b26}&rank=1&20ms&753ms\\
    Ok-topk \cite{b13}&k=1\%&500ms&674ms\\
  \hline
\end{tabular}}
\end{table}

\section{Preliminary}

\subsection{Data Parallel Training}

In default DP training, each worker launches collective operations (e.g., \verb|AllReduce|) immediately to send their local gradients after the computations of all gradients are finished. The total training time for one iteration in DP becomes:
\setlength\abovedisplayskip{1pt}
\setlength\belowdisplayskip{1pt}
\begin{equation}
  T_{DP}=T_{before}+T_{comp}+T_{comm}
\end{equation}
where $T_{before}$ is the time duration before the backward pass, including data input and forward pass. $T_{comp}$ is the computation time of backward pass. $T_{comm}$ is the communication overhead. In this paper, we focus on homogeneous clusters, so $T_{DP}$ is the same for every worker.

The multi-node training time of optimal DP (i.e., time of linear scaling DP) denoted by $T_{DP-LS}$ is equivalent to single-node training time without communications (i.e., $T_{comm}=0$ in (1)). Then, the speedup of DP (compared with training locally using a single device) can be represented as: $P*T_{DP-LS}⁄T_{DP}$, where $P$ is the number of workers. The speedup of linear scaling is equal to $P$, which is an upper limit.

Let the communication-to-computation ratio ($CCR$) be the ratio of communication overhead and computation time (i.e., $T_{comm}=CCR*T_{comp}$). Since $CCR$ is generally greater than 1 because of the communication bottleneck, the speedup of DP can be expressed as:
\setlength\abovedisplayskip{1pt}
\setlength\belowdisplayskip{1pt}
\begin{equation}
  P*\frac{T_{DP-LS}}{T_{DP}}=P*\frac{\frac{T_{before}}{T_{comp}}+1}{\frac{T_{before}}{T_{comp}}+1+CCR}
\end{equation}

\indent Since $T_{before}$ and $T_{comp}$ only depend on computation times, we can easily replace $\frac{T_{before}}{T_{compute}} +1$ with k, then the speedup of DP becomes: $P*\frac{k}{k+CCR}$ , which illustrates that once $CCR$ is high (i.e., communication bottleneck), the performance of DP will be far away from linear scaling. Note that $CCR$ depends on many factors of training tasks, including the type of DNN, training hyperparameters, network environment (bandwidth) and number of workers, etc. Table \uppercase\expandafter{\romannumeral1} shows the $CCR$s of different training tasks in our environment.

\subsection{Overlapping}
Overlapping aims to parallelize computations with communications in DP. Wait-free Back-propagation \cite{b21} is the theoretical basis of Overlapping. For more efficient use of the network bandwidth, merging small tensors of one layer's gradients to a large tensor including several layers’ gradients is introduced to implement better Overlapping \cite{b2,b14,b20}. Once a tensor's gradient is calculated, the communication operation is called on the entire tensor. Equation (3) shows the total training time for one iteration of DP in tensor-based Overlapping:
\setlength\abovedisplayskip{1pt}
\setlength\belowdisplayskip{1pt}
\begin{equation}
\begin{aligned}
  T_{ovlp}= &T_{before}+T_{comp}^{1}+T_{comm}^{b}+\\&max\{\sum_{i=2}^{b}T_{comp}^{i},\sum_{i=1}^{b-1}(T_{comm}^{i}+{bubble}^{i})\}
\end{aligned}
\end{equation}
where b is the number of tensors, $T_{compute}^i$ and $T_{comm}^i$ are the computation time and communication overhead of the $i_{th}$ tensor. In case of shorter communication in DP, we introduce the concept bubble in (3), where ${bubble}^i$ represents the idle time of communication when the computation time of the ${i+1}_{th}$ tensor is longer than the communication time of the $i_{th}$ tensor, as shown in Fig. 1(d). Since the network speed is relatively slow in most cases (i.e., $CCR$$\geq$1), communication operations usually happen back to back, and the ${bubble}^i$ will not appear.

In the more common case where computation time is less than communication time, (3) can be simplified to:
\setlength\abovedisplayskip{1pt}
\setlength\belowdisplayskip{1pt}
\begin{equation}
  T_{ovlp}=T_{before}+\sum_{i=1}^{n}T_{comp}^i+T_{comm}^{'}
\end{equation}
where $T_{comm}^{'}$ represents the time of the part of communication that cannot be overlapped with computation, as Fig. 1(b) shows. Since the communication overhead equals to $CCR*\sum_{i=1}^{n}T_{comp}^i$ and Overlapping can parallelize the communication overhead at most equal to the computation time, $T_{comm}^{'}$ can be approximated as $(CCR-1)*\sum_{i=1}^{n}T_{comp}^i$ , which means once $CCR$ is much higher than 1, $T_{comm}^{'}$ will be large and $T_{ovlp}$ will be much longer than $T_{DP-LS}$. The experiments in Table \uppercase\expandafter{\romannumeral1} further confirmed that the speedups provided by Overlapping (denoted by $S_{ovlp}$) are negatively correlated to $CCR$ and much less than the speedups of linear scaling.

\subsection{Gradient Compression (GC)}
Existent GC schemes often suffer from non-negligible compression overhead. When using GC, the training time for one iteration in DP becomes:
\setlength\abovedisplayskip{1pt}
\setlength\belowdisplayskip{1pt}
\begin{equation}
  T_{GC}=T_{before}+T_{comp}+T_{compress}+T_{comm-GC}
\end{equation}
where $T_{compress}$ is the compression overhead (including compression and decompression), and $T_{comm-GC}$ is the communication overhead compressed by GC. In Table \uppercase\expandafter{\romannumeral2}, we observe that the DP training time reductions in some GC schemes are impaired by their high compression overhead. For example, although Top-k reduces communication overhead by about 600ms in each iteration of training VGG-19, it has a significant compression overhead of 1560ms. In contrast, PowerSGD costs less training time by 733ms with respect to communication overhead reduction by about 753ms and only 20ms compression overhead. Therefore, training the same DNN using PowerSGD is several times faster than Top-k.

For additional training time reduction, we apply Overlapping into GC with the training time of one iteration denoted by (6), which is derived from (4) and (5):
\setlength\abovedisplayskip{1pt}
\setlength\belowdisplayskip{1pt}
\begin{equation}
  T_{GC\&ovlp}=T_{before}+\sum_{i=1}^{n}(T_{comp}^i+T_{compress}^i)+T_{comm-GC}^{'}
\end{equation}
where $T_{compress}^i$ is the compression overhead of the $i_{th}$ tensor, $T_{comm-GC}^{'}$ represents the time of the part of compressed communication that cannot be overlapped with computation. As discussed in Section \uppercase\expandafter{\romannumeral2}.B, $T_{comm-GC}^{'}$ can be approximated as $(CCR-1)*\sum_{i=1}^n(T_{comp}^i+T_{compress}^i)$. Since using GC results in much lower $CCR$, $T_{comm-GC}^{'}$ is much smaller and even near zero. Unfortunately, data dependency in some GC schemes \cite{b11,b13,b16,b26} makes (6) not available. We choose 2 GC schemes with no data dependency: Random-k and FP16, to apply GC and Overlapping concurrently. Results in Table \uppercase\expandafter{\romannumeral3} illustrate that reducing the $CCR$ of DP to around 1 by GC achieves near-linear scaling for DP training with Overlapping. 

\begin{table}
\setlength{\abovecaptionskip}{0.cm}
\setlength{\belowcaptionskip}{-0.cm}
\setlength{\leftskip}{-2pt}
  \caption{Examples of using GC and Overlapping concurrently}
  \resizebox{86.5mm}{5mm}{
  \begin{tabular}{cccccc}
    \hline
    GC schemes&$CCR$&$CCR$ after compression&$S_{GC}$&$S_{GC-ovlp}$&$S_{LS}$\\
    \hline
    Random-k&\multirow{2}{*}{2.1} &1.07&1.29x&2.05x&\multirow{2}{*}{2.67x}\\
    FP16& &1.04&1.42x&2.35x&\\
  \hline
\end{tabular}}
\end{table}

\subsection{Opportunities and Challenges}
From Section \uppercase\expandafter{\romannumeral2}.B, we know that the high $CCR$ of DP makes Overlapping difficult to achieve linear scaling. Although GC is an effective technique to reduce $CCR$, applying GC into Overlapping raises new challenges: high compression overhead and additional data dependency. Compression overhead generally comes from high time complexity operators in GC schemes, and there is a trade-off between compression overhead and training accuracy when compression ratio is high. However, we observe that in common network with 30-100Gbps bandwidth, the $CCR$s of common DP tasks are not too high, as shown in Table \uppercase\expandafter{\romannumeral1}. Therefore, it is possible for us to redesign a novel GC scheme with lower compression overhead.

Table \uppercase\expandafter{\romannumeral1} also shows that $CCR$ may differ significantly in different DP tasks. To achieve linear scaling, using a constant compression ratio may not be applicable to all DP tasks. $CCR$ depends on many factors, such as network bandwidth, DNN type, communication library, and worker number. It is also a challenge to obtain accurate $CCR$ with cost-tolerant profiling.

\section{Methodology}

In this section, we first introduce our new GC scheme with a coarse-grained gradient filter to achieve near-zero compression overhead. We then present our automatic compression ratio selection strategy based on a distributed profiler. Next, we design a tensor sharding technique to alleviate the communication bottleneck on large tensors. Finally, we introduce how error feedback is employed to preserve training accuracy and give a convergence analysis.

\begin{figure}[t]
\setlength{\abovecaptionskip}{0.cm}
\setlength{\belowcaptionskip}{-0.cm}
  \centering
  \includegraphics[width=86.5mm]{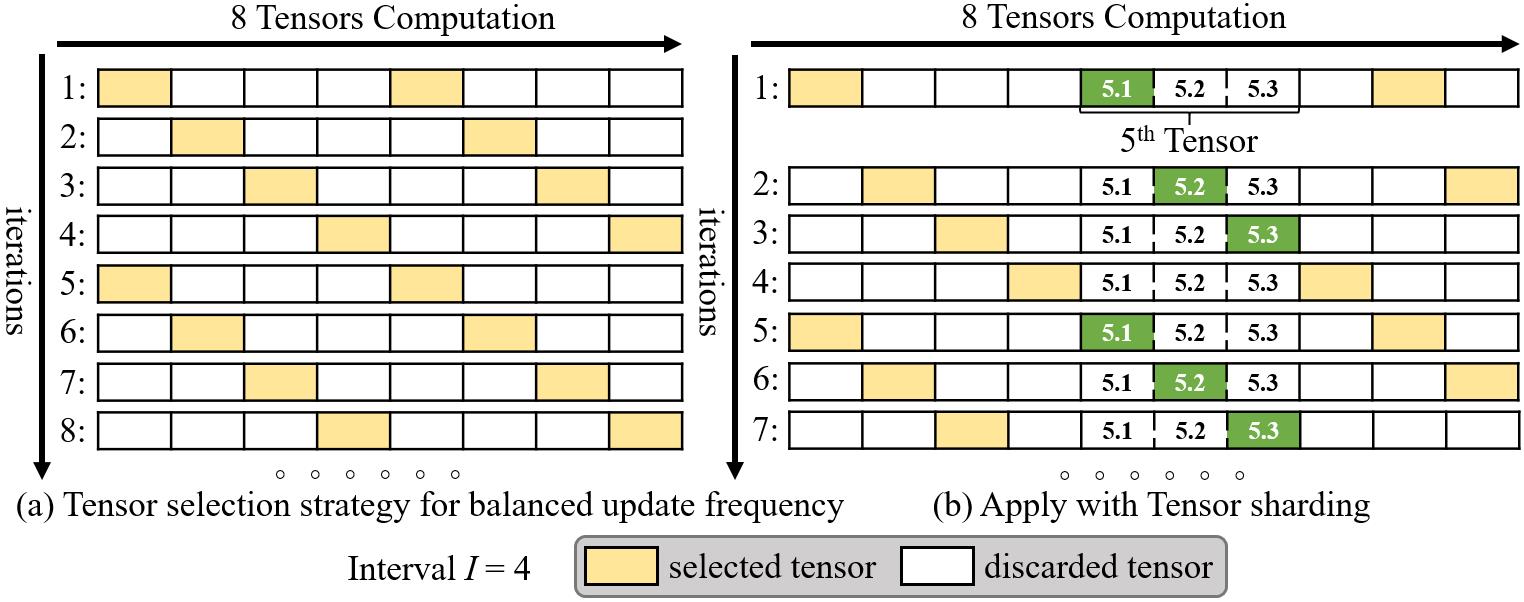}
  \caption{Examples of COVAP’s tensor selection strategy.}
\end{figure}

\subsection{Coarse-grained Gradient Filter}
In Gradient Compression, the filter represents the gradient selection strategy and often appears in sparsification schemes. Existing fine-grained filters (e.g., Top-k) select gradients according to each iteration’s overall gradient value distribution. However, such strategies often suffer from non-negligible compression overhead, as shown in Table \uppercase\expandafter{\romannumeral2}. Instead, we design a novel coarse-grained gradient filter to decrease time complexity. COVAP considers the implementations of DL frameworks and uses communication unit \verb|tensor| as the granularity of filters. In practice, our filter discards the complete communication operations of some \verb|tensor|s in each DP iteration.

COVAP’s coarse-grained filter can significantly reduce the time complexity. Tensors for communication often include several consecutive layers of DNN, which implies that the number of communication tensors is much lower than the number of gradients. For example, the default communication tensor size in PyTorch is 25MB. For DNNs of 50 million parameters with a size of about 200MB, PyTorch allocates about 8 communication tensors to contain all gradients (here, we assume the size of each layer is balanced). Since COVAP’s filter only needs to select tensors from those 8 tensors instead of traversing all 50 million gradients, the time complexity is significantly reduced. Another benefit is that from the implementation perspective, using the same granularity of communication operations with DL frameworks will not introduce any additional overhead of rebuilding \verb|tensor|s in each iteration. Empirically, the \verb|tensor| size is set to 25MB, the same as the default value in Pytorch. Less or greater sizes may result in performance degradation, as presented in \cite{b14}.  

In such coarse-grained filter, the tensor selection strategy is also different from other GC schemes. In each iteration, each worker chooses tensors with same index to communicate with others. Tensors are selected alternately, which means each tensor is only communicated once in every $I$ iterations ($I$ represents the interval). Meanwhile, only one tensor is chosen for communication from every $I$ tensors in each iteration. Tensor $t$ is selected in iteration $num\_steps$ when $(t+num\_steps)\%I=0$. Fig. 2(a) illustrates the above selection strategy. When the interval $I$ is 4, the first tensor is communicated at the first and $5^{th}$ iterations, the second tensor is communicated at the second and $6^{th}$ iterations, and so on. One benefit of such a strategy is that each tensor is communicated and updated by the same interval, which can alleviate the negative effect of staleness \cite{b30} on training convergence. Another benefit is that it does not introduce any additional data dependencies mentioned in Section \uppercase\expandafter{\romannumeral1} because each worker can select tensors by themselves according to the current iteration numbers and the interval (The interval is determined in the early stage of training and remains unchanged in subsequent training, we give a detailed discussion in section \uppercase\expandafter{\romannumeral3}.B.), which means no additional communication is required to synchronize the tensor selection results.

\begin{figure}[t]
\setlength{\abovecaptionskip}{0.cm}
\setlength{\belowcaptionskip}{-0.cm}
  \centering
  \includegraphics[width=86.5mm]{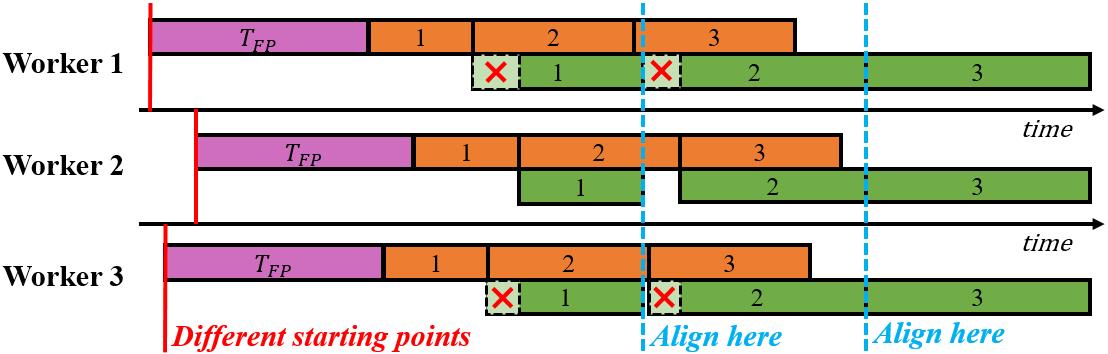}
  \caption{The example visualization of distributed profiler. The timelines of all processes align at the dotted lines. The light green dotted squares represent the idle time of worker 1 and 3 in communication waiting for worker 2.}
\end{figure}

\subsection{Compression ratio selection}
The coarse-grained filter has minimized compression overhead. However, for various training tasks in different computing environments, a constant compression ratio is difficult to achieve fully overlapping consistently. In other words, $CCR$s are different in different cases. For example, using different GPU types can significantly influence the $CCR$: replacing the GPU from V100 to A100 will speed up the computation and increase $CCR$. Therefore, to achieve full overlapping in most cases, COVAP adopts different compression ratios for different cases. Note that in this paper, we do not consider the interference from other tasks in the cluster or other performance fluctuations, which means $CCR$ will not change rapidly throughout whole training. 

From the analysis in Section \uppercase\expandafter{\romannumeral2}.D, COVAP should decrease communication time by $CCR$ times. Since COVAP uses the same collective communication primitives (i.e., \verb|AllReduce|) as the baseline of no compression, decreasing communication overhead by $CCR$ times is equivalent to reducing communication data volume by $CCR$ times. Because communication tensors of DL frameworks are generally in fixed sizes \cite{b14,b20}, COVAP initially assumes each tensor has the same size. Then, reducing data volume by $CCR$ times can be approximated as reducing the number of communicating tensors by $CCR$ times. We can infer from the algorithm in Section \uppercase\expandafter{\romannumeral3}.A that the interval $I$ in our scheme should equal $CCR$.

COVAP measures the $CCR$ in the early stage of the training task and sets its compression ratio according to the result. For implementation details, COVAP measures the $CCR$s of different DNNs based on the PyTorch profiler module. PyTorch profiler’s context manager API can be used to study device kernel activity and visualize the execution trace. Specifically, we use the \verb|profiler_cuda| module to track \verb|cuda| events, including communication (i.e., \verb|AllReduce| and \verb|AllGather|) and computation (i.e., forward and backward propagation) operators. Since we only need to profile one training iteration to obtain the communication and computation time, COVAP only costs a one-off profiling overhead (less than 5 seconds in our experiments). Compared to hours of training, such profiling overhead is acceptable.

The profiler tool provided by PyTorch is applicable for stand-alone programs. In distributed training, PyTorch creates multi-processes for different workers. Although we can use the original profiler to monitor one of the processes and obtain its exact time of computation events in one iteration, the communication time measured in this way may be inaccurate. That is because there may be a slight time difference in distributed training when different workers start their communications on different nodes, as Fig. 3 shows. In such cases, the communication operators of some workers who finish their computations early will be longer than others since they must wait for other workers to rendezvous in communication. In our experiments, such waiting may cause a 20\% communication time measurement error. Therefore, we develop a distributed profiler. The key point is to align the timeline at the end of each communication operator in each training step to eliminate measurement errors caused by such communication waiting latency.

Using that distributed profiler, we can obtain more accurate communication time to compute $CCR$. Then, COVAP sets the interval $I$ of the filter according to that $CCR$. Since $I$ must be an integer but measured $CCR$s may not be integers, we let $I$ equals to $\lceil CCR \rceil$, which implies that COVAP compresses communication by a little more than $CCR$ times to ensure as much communication as possible can be overlapped with computation in DP.

\subsection{Tensor Sharding}
In Section \uppercase\expandafter{\romannumeral3}.B, we assumed that each communication tensor has the same size as the default fixed size of DL frameworks. However, the practical tensor sizes may vary significantly according to severe imbalanced sizes of DNN layers, which could cause communication bottlenecks if the large tensors are selected to communicate in COVAP. In practice, DL frameworks allocate gradients of layers into tensors. The gradient tensor of one layer is used as the minimum unit in allocations, which means each tensor contains integral number of layers and at least one. Even if a layer’s size is much larger than the tensor size of the framework, PyTorch will not split such a large variable into multiple tensors. Although DL frameworks often adopt a much larger default tensor size than the size of most layers in DNNs, there may still exist some much larger layers than the default tensor size in some DNNs. For example, we list some of the layers of VGG-19 in Table \uppercase\expandafter{\romannumeral4}. It can be seen that the parameter number of Layer FC1 is much larger than other layers, accounting for 71.53\% of all parameters. The size of Layer FC1 is 102760448 * 4 Bytes = 401.4 MB. Compared with the default tensor size of 25MB in PyTorch DDP, the tensor of that layer is oversized. We also tested the communication time of all tensors when training VGG-19 across 8 nodes, and the result is shown in Table \uppercase\expandafter{\romannumeral5}. The communication of that large tensor costs about 603.238ms, accounting for 72.67\% of total communication time. Such prominent communications of large layers may impair the Overlapping advantage in COVAP due to the imbalance, as shown in Fig. 4(b).

\begin{table}\small
\setlength{\abovecaptionskip}{0.cm}
\setlength{\belowcaptionskip}{-0.cm}
\setlength{\leftskip}{-2pt}
  \caption{Layer sizes of VGG-19}
  \centering
  \label{tab:vgg}
  \resizebox{55mm}{22mm}{
  \begin{tabular}{ccc}
    \hline
    Layer name&parameters&ratio\\
    \hline
    Input&---&---\\
    
    Conv1\_1&1728&0.00\%\\
    
    Conv1\_2&36864&0.03\%\\
    
    Pool1&---&---\\
    \multicolumn{3}{c}{......}\\
    Pool5&---&---\\
    
    $\mathbf{FC1}$&$\mathbf{102760448}$&$\mathbf{71.53\%}$\\
    
    $\mathbf{FC2}$&$\mathbf{16777216}$&$\mathbf{11.68\%}$\\
    
    FC3&4096000&2.85\%\\
    
    Softmax&---&---\\
    \hline
    total&143652544&100.00\%\\
  \hline
\end{tabular}}
\end{table}

\begin{table}\small
\setlength{\abovecaptionskip}{0.cm}
\setlength{\belowcaptionskip}{-0.cm}
\setlength{\leftskip}{-2pt}
  \caption{Communication times of tensors in VGG-19}
  \label{tab:vggbucket}
  \resizebox{86.5mm}{15mm}{
  \begin{tabular}{cccc}
    \hline
    Tensor id&Number of elements&comunication time&ratio\\
    \hline
    1&4101096&16.177ms&1.95\%\\
   
    $\mathbf{2}$&$\mathbf{16781312}$&$\mathbf{99.205ms}$&$\mathbf{11.95\%}$\\
    
    $\mathbf{3}$&$\mathbf{107480576}$&$\mathbf{603.238ms}$&$\mathbf{72.67\%}$\\
    
    4&7079424&36.513ms&4.40\%\\
    
    5&7669760&40.743ms&4.91\%\\
    
    6&555072&34.218ms&4.12\%\\
    \hline
    total&143667240&830.094ms&100.00\%\\
  \hline
\end{tabular}}
\end{table}

Therefore, COVAP designs a tensor sharding strategy that slices the large tensor into small pieces such that all tensors are balanced in size. Specifically, after building tensors, COVAP counts the elements’ number of all tensors to find the median number of elements in one tensor. Suppose a tensor has multiple  times elements than that median number. In that case, COVAP evenly slices it into $\lfloor \frac{numel}{median} \rfloor$ parts, where $numel$ represents the elements’ number of that tensor (i.e., the total number of gradients of all layers in that tensor). Note that if $\lfloor \frac{numel}{median} \rfloor$ is larger than interval $I$, COVAP only slices that tensor into $I$ parts for balance.

Then, when COVAP selects tensors to communicate, such a large tensor is regarded as $\lfloor \frac{numel}{median} \rfloor$ tensors, from which COVAP may choose one tensor to communicate in one iteration. Still using VGG-19 as an example, as shown in Table \uppercase\expandafter{\romannumeral5}, the median element number of tensors is 5590260. Thus, the second and third tensors are sliced into 3 and 19 new tensors (if the interval $I$ is less than 3 or 19, COVAP slices them into $I$ parts instead). Then, the total number of tensors is regarded as 26 when COVAP chooses tensors for communication, which results in a much better balance of tensor sizes. Another example of how COVAP selects tensors when using tensor sharding is shown in Fig. 2(b), and how tensor sharding balances the communication time is shown in Fig. 4(c).

\begin{figure}[t]
\setlength{\abovecaptionskip}{0.cm}
\setlength{\belowcaptionskip}{-0.cm}
  \centering
  \includegraphics[width=86.5mm]{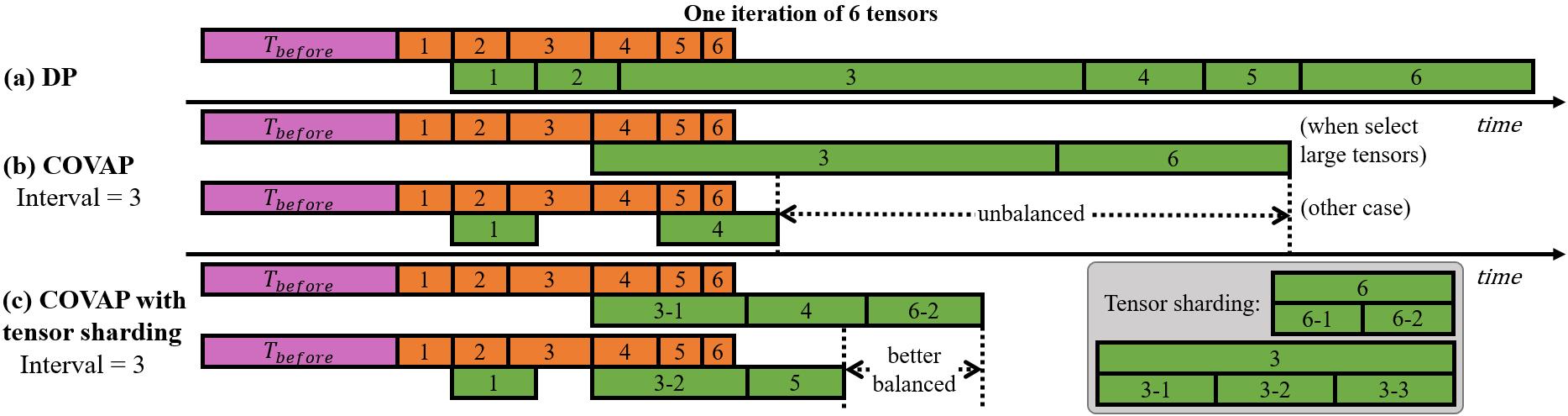}
  \caption{Examples of using tensor sharding for large tensors. In (a), Tensor 3 and 6 are two large tensors. In (b), when COVAP selects Tensor 3 and 6, the training time is much longer than in other cases. In (c), using tensor sharding resulted in a more balanced training time.}
\end{figure}

\subsection{Convergence}
{
\setlength{\parindent}{0cm}
$\mathbf{The}$ $\mathbf{error}$ $\mathbf{feedback}$ $\mathbf{scheduler}$. Although COVAP significantly improves training speed with techniques in Section \uppercase\expandafter{\romannumeral3}.A, \uppercase\expandafter{\romannumeral3}.B and \uppercase\expandafter{\romannumeral3}.C, such compression schemes are theoretically lossy in training accuracy. To alleviate such a problem, COVAP equips error feedback (i.e., local memory accumulation) to ensure convergence and designs a new scheduler to continuously adjust a compensation coefficient of Error feedback in training, which balances the feedback amount of errors. 
}

{
\vspace{0.2cm}
\setlength{\parindent}{-0.1cm}
\setlength{\leftskip}{-2pt}
\resizebox{87.5mm}{15mm}{
  \begin{tabular}{lll}
    \hline
    \multicolumn{3}{l}{$\mathbf{Algorithm\ 1}$ Error feedback in compression given gradient $G$.}\\
    \hline
    1:& &$\mathbf{function}\ compression(G)$\\
    2:& & $\ $ $G$ += $residuals$\\
    3:& & $\ $ $G^{’}$ = sparsification\_or\_quantization($G$)\\
    4:& & $\ $ $residuals$ = $G$ - $G^{’}$\\
    5:& & $\ $ return $G^{’}$\\
    6:& &$\mathbf{end\ function}$\\
  \hline
\end{tabular}
}
}

The idea of error feedback was first introduced in one-bit \cite{b19}. Following GC works \cite{b3,b7,b11,b13,b16,b24,b26} widely adopted error feedback as a compensation strategy to ensure convergence and preserve accuracy. Several works \cite{b23,b24} also conducted theoretical analyses. Generally, error feedback is used as shown in Algorithm 1, where $residuals$ represent the D-values saved in local memory.

{
In COVAP, we only choose a few tensors to communicate in each iteration. Applying error feedback to COVAP means we save other tensors’ gradients in local memory and add them to their corresponding tensor in the next iteration, just like Algorithm 1. However, in practice, we found that using error feedback in such a way may lead to divergence due to the harmful effect of the staleness of delayed update parameters. To alleviate the staleness effect, we introduce the error feedback scheduler into Algorithm 1 based on the observation in \cite{b10} that a large compensation coefficient in early training epochs may harm model accuracy. 
}

The error feedback scheduler multiplies residuals by a compensation coefficient before the residuals are added to current gradients (i.e., line 2 in Algorithm 1). Similar to the learning rate scheduler of variant declining strategies, our scheduler adjusts that coefficient continuously with increasing strategy in training. The coefficient is equal to:
\begin{displaymath}
  min\{init\_value + \lfloor\frac{num\_steps}{ascend\_steps}\rfloor*ascend\_range,1\},
\end{displaymath}
where $init\_value$ represents the initial value of the coefficient, $num\_steps$ represents the number of iterations trained, $ascend\_steps$ represents the interval between ascending, and $ascend\_range$ represents the ascending range. The max value of the coefficient is 1.

{
\setlength{\parindent}{0cm}
$\mathbf{Convergence}$ $\mathbf{analysis}$. From Section \uppercase\expandafter{\romannumeral3}.A, the filter of COVAP is similar to sparsification schemes since COVAP also selects a subset of all gradients in each iteration. Similar to Top-k\cite{b3}, COVAP can be regarded as a particular case of Random-k \cite{b23}. Therefore, we can follow the theoretical proof in \cite{b23}.
}

We first give the definition of COVAP's compression operator:\\
{
\setlength{\parindent}{0cm}
$\mathbf{Definition}$ $\mathbf{1.}$ $For$ $1\leq l\leq h,$ $where$ $h$ $is$ $the$ $depth$ $of$ $network$ $\mathbf{x},$ $the$ $compression$ $operator$ $COVAP$ $is$ $defined$ $for$ $\mathbf{x}\in \mathbb{R}^d$ $as$
\begin{small}
\begin{displaymath}
    (COVAP(\mathbf{x}))_l := 
    \begin{cases}
        (\mathbf{x})_l,& if\enspace(l+num\_steps)\equiv 0(mod\enspace I),\\
        0, & otherwise,
    \end{cases}
\end{displaymath}
\end{small}
$where$ $num\_steps$ $is$ $current$ $number$ $of$ $training$ $iterations$ $and$ $I$ $is$ $the$ $interval$ $set$ $by$ $COVAP.$
}

Same as Random-k and Top-k, operator $COVAP$ also satisfies the definition of being a k-contraction in \cite{b23}:
\begin{displaymath}
    \mathbb{E}||\mathbf{x}-COVAP(\mathbf{x})||^2 \leq (1-\frac{k}{d})||\mathbf{x}||^2, \quad \forall \mathbf{x} \in \mathbb{R}^d
\end{displaymath}
Then, we utilize the convergence proof process for Top-k and Random-k in the non-convex case presented in the work of \cite{b23} to prove the convergence of COVAP.

\section{Evaluation}
\subsection{Experimental setup}

\begin{figure*}[t]
\setlength{\abovecaptionskip}{0.cm}
\setlength{\belowcaptionskip}{-0.cm}
  \centering
  \includegraphics[width=\textwidth]{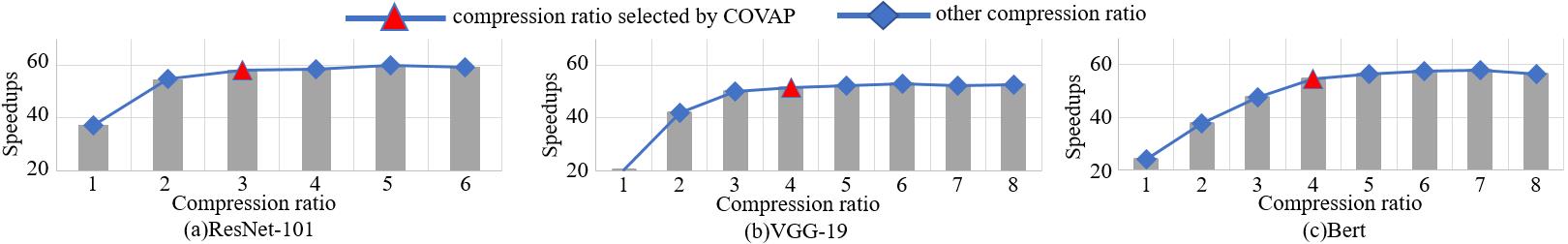}
  \caption{The speedups provided by COVAP with different compression ratios when training different DNNs on 64 GPUs.}
\end{figure*}

\begin{figure*}[t]
\setlength{\abovecaptionskip}{0.cm}
\setlength{\belowcaptionskip}{-0.cm}
  \centering
  \includegraphics[width=\textwidth]{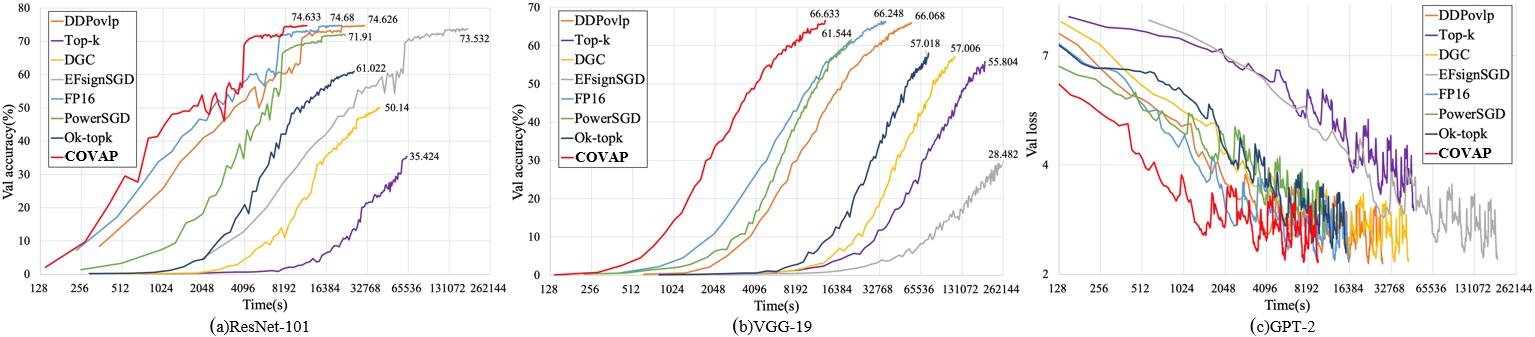}
  \caption{Time-to-solution curves of different GC schemes when training 3 DNNs.}
\end{figure*}

We conducted our experiments on 8 Alibaba Cloud ECS instances. Each instance contains an Intel Xeon Platinum 8163 CPU and 8 NVIDIA V100 GPUs with 16 GB global memory. We choose a typical bandwidth of 30Gbps in the public cloud. In High-Performance Computing, the bandwidth is usually much higher and reaches 100Gbps. Due to the adaptive compression ratio relying on $CCR$, we believe COVAP can also provide acceleration in such scenarios.

\begin{table}
\setlength{\abovecaptionskip}{0.cm}
\setlength{\belowcaptionskip}{-0.cm}
\setlength{\leftskip}{-2pt}
  \caption{Neural networks used for evaluation}
  \label{tab:evaluation}
  \resizebox{86.5mm}{9mm}{
  \begin{tabular}{cccc}
    \hline
    Task&DNN&parameters&datasets\\
    \hline
    \multirow{2}{*}{Image classification}&ResNet-101&44654504&\multirow{2}{*}{Cifar-10/ImageNet}\\
    &VGG-19&143652544&\\
    Text classification&Bert&102267648&\multirow{2}{*}{THUC-News}\\
    Text generation&GPT-2&81894144&\\
  \hline
\end{tabular}}
\end{table}

\begin{table*}
\setlength{\abovecaptionskip}{0.cm}
\setlength{\belowcaptionskip}{-0.cm}
\setlength{\leftskip}{-2pt}
  \caption{training times and accuracies of different DNNs}
  \label{tab:casestudy}
  \resizebox{\textwidth}{14mm}{
  \begin{tabular}{cccccccccccccccc}
    \hline
    \multirow{2}{*}{GC schemes}&\multicolumn{3}{c}{ResNet-101}& &\multicolumn{3}{c}{VGG-19}& &\multicolumn{3}{c}{Bert}& &\multicolumn{3}{c}{GPT-2}\\
    \cline{2-4}\cline{6-8}\cline{10-12}\cline{14-16}
    &time(s)&accuracy(\%)&speedup& &time(s)&accuracy(\%)&speedup& &time(s)&accuracy(\%)&speedup& &time(s)&loss&speedup \\
    \hline
    DDPovlp&31260.4&74.626&21.52&&56201.9&66.068&12.07&&729.8&94.58&26.64&&28296.9&1.922&19.10\\
    Top-k&64158.6&35.424&10.48&&192920.2&55.804&3.51&&1271.4&94.46&15.28&&47570.6&2.697&11.366\\
    DGC&40516.6&50.14&16.64&&115819.0&57.006&5.85&&431.2&94.33&45.12&&43772.2&1.953&12.35\\
    EFsignSGD&179978.9&73.532&3.76&&255388.2&28.482&2.66&&978.5&93.88&19.92&&192404.4&1.974&2.81\\
    FP16&21312.3&74.68&31.6&&36293.4&66.248&18.68&&473&94.38&41.12&&16167.6&1.945&33.44\\
    PowerSGD&22485.2&71.916&29.92&&20320.5&61.544&33.37&&428.9&93.98&45.36&&10844.9&2.253&49.85\\
    Ok-topk&26292.6&61.022&25.6&&74646.3&57.018&9.08&&609.6&93.97&31.92&&15390.2&2.061&35.13\\
    COVAP&11696.9&74.633&57.52&&13090.4&66.633&51.80&&336.4&94.44&57.84&&9635.4&1.937&56.11\\
  \hline
\end{tabular}}
\end{table*}

We use four neural networks from different deep learning domains summarized in Table \uppercase\expandafter{\romannumeral6} for evaluation. For VGG-19 and ResNet-101, we use SGD optimizer with an initial learning rate of 1e-3; for Bert and GPT-2, we use Adam optimizer with initial an learning rate of 5e-5 and 1.5e-4. We compare our scheme COVAP with the state-of-the-art and other popular GC schemes in Table \uppercase\expandafter{\romannumeral2}. For a fair comparison, all schemes are implemented by the DDP communication hook in PyTorch 1.9.0 using \verb|NCCL| as the communication library.

Most GC implementations are referred to \cite{b28}. Note that Ok-topk uses \verb|mpi4py| as its communication library. We reimplemented it in our environment, which may result in a different performance than reported in \cite{b13}.

\subsection{Compression ratio selection}
We verify the correctness of the adaptive compression ratio selection strategy introduced in Section \uppercase\expandafter{\romannumeral3}.B on three different DNNs. Fig. 5 shows the performance of COVAP in using different compression ratios. We tested the speedups provided by COVAP in all cases. The speedups were calculated by (1) and (2) such that the upper limit speedup of DP training with 64 GPUs is 64, as dotted lines shown in Fig. 5. Compression ratio 1 represents the baseline without compression (i.e., default PyTorch DDP).

As discussed in Section \uppercase\expandafter{\romannumeral3}.B, COVAP chooses $\lceil CCR\rceil$ as its compression ratio. Using a higher compression ratio than $\lceil CCR\rceil$ is meaningless because COVAP has already reduced most communication overhead through Overlapping. In Fig. 5(a), the speedup of training ResNet-101 increased very slowly as the compression ratio was beyond 3. Similarly, Fig. 5(b) and 5(c) discover the maximum speedups were 51.51 and 54.55 when the compression ratio was 4, which COVAP selected.

\subsection{Case studies on training time and convergence}
We studied the training time and model convergence using real-world applications listed in Table \uppercase\expandafter{\romannumeral6}. In addition to listing the results of each complete training, we make a further breakdown of the training time of one iteration for better understanding, including compression (e.g., Top-k selection from the gradients), communication (e.g., \verb|AllReduce|), and computation (i.e., forward and backward pass). The experiments were conducted in an 8-node cluster with a 30Gbps network, each node with 8 NVIDIA V100 GPUs. Note that we use average values for breakdowns since some GC schemes’ training speed varies significantly at different stages of training. For linear scaling, the speedup should be 64 (compared with one GPU), which is also an upper limit of all schemes. Besides, since Random-k diverged in most experiments, we do not give the convergence result of Random-k. Table \uppercase\expandafter{\romannumeral7} presents the training time and accuracy results of different GC schemes when training 4 DNNs, wherein the speedup is calculated by (2) in Section \uppercase\expandafter{\romannumeral2}.A. Fig. 6 presents the time-to-solution curves of 3 DNNs except for Bert, since we only use the title of each news in THUC-News for text classification and its training time is relatively short. DDPovlp represents the default PyTorch with Overlapping and is the baseline with no gradient compression.

\begin{figure}[t]
\setlength{\abovecaptionskip}{0.cm}
\setlength{\belowcaptionskip}{-0.cm}
  \centering
  \includegraphics[width=86.5mm]{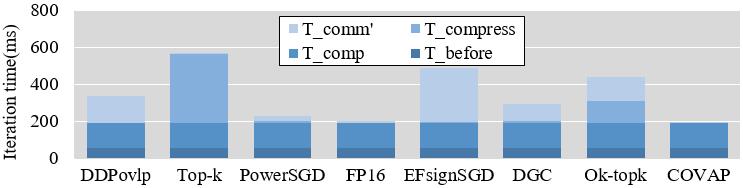}
  \caption{Breakdown of iteration time in ResNet-101.}
\end{figure}

\begin{figure}[t]
\setlength{\abovecaptionskip}{0.cm}
\setlength{\belowcaptionskip}{-0.cm}
  \centering
  \includegraphics[width=86.5mm]{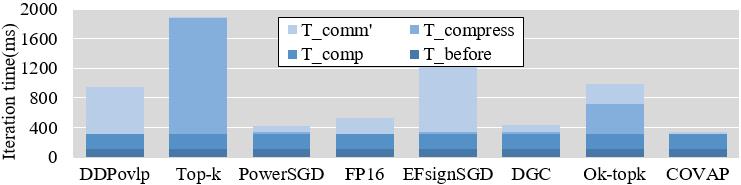}
  \caption{Breakdown of iteration time in VGG-19.}
\end{figure}

\subsubsection{ResNet-101}
Fig. 7 shows the breakdown of each iteration’s training time of different GC schemes when training ResNet-101 on ImageNet. $T\_comm$' represents communication overhead that cannot be overlapped with computation. Although Top-k hid almost all communication overhead ($T\_comm$'$ \approx 0$), it had a considerable compression overhead of about 370ms, which resulted in 2.05x longer training time than the baseline. Besides, since some schemes use synchronized communications, they are incompatible with Overlapping. One example is Ok-topk. Although it has a lower compression overhead of about 40ms than 370ms of Top-k, its communication cannot be overlapped with computation, so its training speed only exceeded the baseline by 1.19x. Another example is PowerSGD. Although it had the lowest communication overhead of 20ms and low compression overhead of 10ms, its performance did not outperform COVAP and FP16. 
From Table \uppercase\expandafter{\romannumeral7} and Fig. 6(a), COVAP achieved the best performance among all GC schemes, outperformed the others by 1.92x-15.39x for the total training time and achieved linear scaling. For model convergence, only 3 GC schemes achieved similar accuracy as the baseline, and all sparsification schemes (including Top-k, DGC, and Ok-topk) did not reach the desired accuracy.

\subsubsection{VGG-19}
Fig. 8 shows the breakdown of VGG-19 training time. Notably, ResNet-101 had a lower $CCR$ of 2.1 compared to VGG-19’s $CCR$ of 4.0 due to the larger parameter amount of VGG-19, which implies training VGG-19 has more severe communication bottleneck so that the baseline DDPovlp had lower speedups from 21.52x on ResNet-101 to 12.07x on VGG-19. In such a case, the speedup of FP16 is only 18.68x compared to 31.6x on ResNet-101 due to insufficient compression. Besides, only COVAP did not suffer from 2x-6.67x larger compression overhead, which benefited from the low time complexity of our coarse-grained filter. Moreover, COVAP adaptively selected its compression ratio to 4 according to the higher $CCR$ of VGG-19. In VGG-19 training, COVAP outperformed all other schemes by 1.55x-19.51x in training time.

Fig. 6(b) shows the time-to-solution curves of different GC schemes when training VGG-19. For convergence, only COVAP and FP16 achieved similar accuracy as the baseline. It is noteworthy that some GC scheme has better (e.g., sparsification) or worse (e.g., EfsignSGD) accuracy when training VGG-19 compared with training ResNet-101, which indicates that those GC schemes may only guarantee convergence and accuracy on specific DNNs.

\begin{figure}[t]
\setlength{\abovecaptionskip}{0.cm}
\setlength{\belowcaptionskip}{-0.cm}
  \centering
  \includegraphics[width=86.5mm]{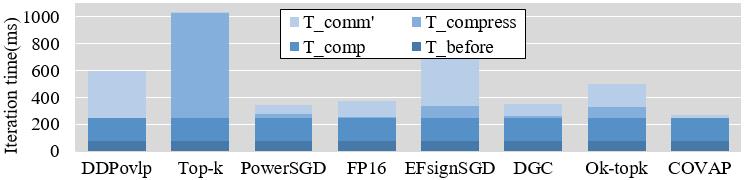}
  \caption{Breakdown of iteration time in Bert.}
\end{figure}

\begin{figure}[t]
\setlength{\abovecaptionskip}{0.cm}
\setlength{\belowcaptionskip}{-0.cm}
  \centering
  \includegraphics[width=86.5mm]{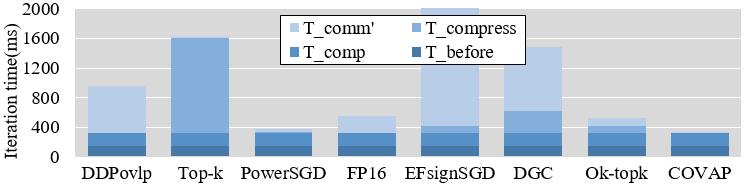}
  \caption{Breakdown of iteration time in GPT-2.}
\end{figure}

\begin{figure*}
\setlength{\abovecaptionskip}{0.cm}
\setlength{\belowcaptionskip}{-0.cm}
  \centering
  \includegraphics[width=\textwidth]{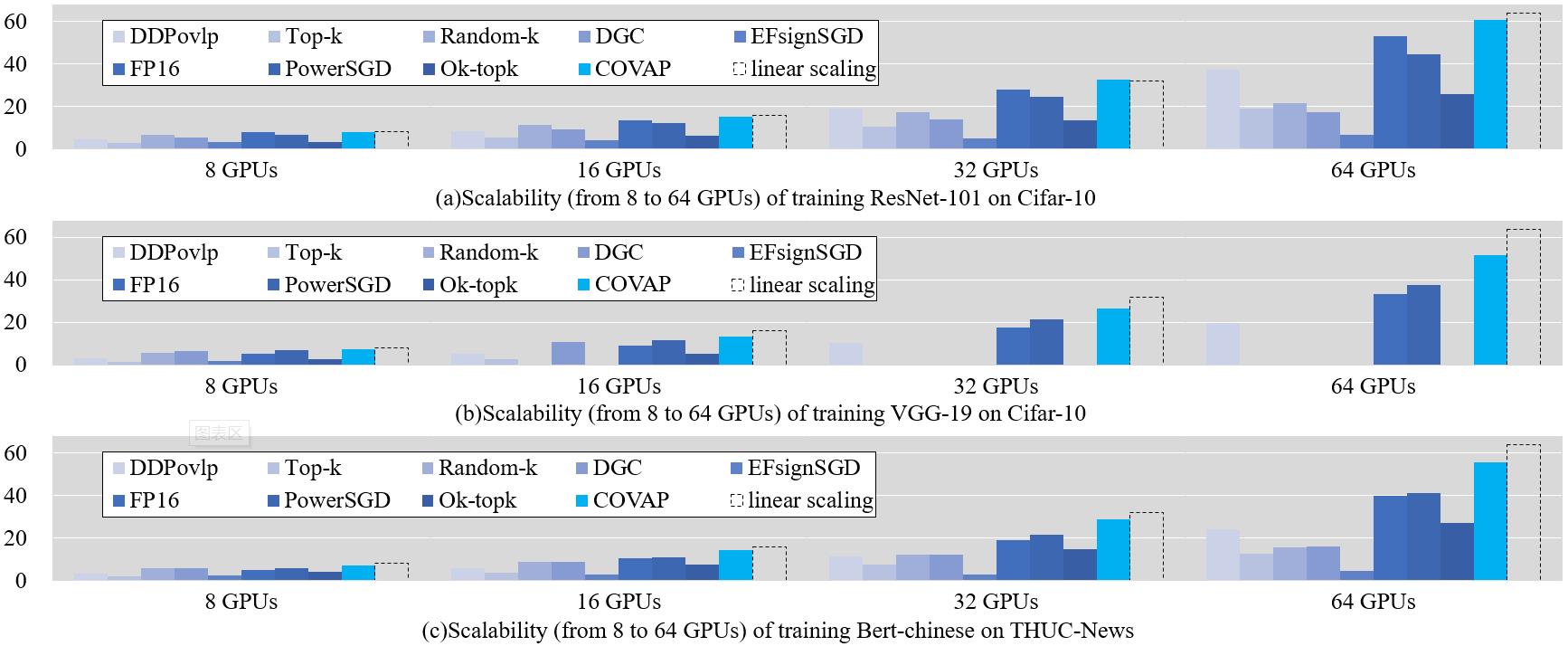}
  \caption{Speedups of different GC schemes under four different cluster sizes.}
\end{figure*}

\subsubsection{Bert}
Fig. 9 shows the breakdown of Bert training time. Similar to the experiments on ResNet-101 and VGG-19, COVAP had a better performance than all counterparts, outperforming them by 1.27x-3.78x in training time. Regarding model convergence, the accuracy achieved by COVAP was 0.14\% lower than the baseline. Among all schemes, COVAP achieved the fastest time-to-solution and near-linear scaling in 9.6\% less speedup than the optimal case.

\subsubsection{GPT-2}
Fig. 10 shows the breakdown of GPT-2 training time. The $CCR$ of GPT-2 measured by our distributed profiler is about 3.5, so we set the compression ratio to 4. Fig. 6(c) shows time-to-solution of all GC schemes. Unlike CV tasks, only Top-k had apparent lower accuracy than the baseline. In GPT-2 training, COVAP outperformed all other schemes by 1.13x-19.96x in training time while achieving expected accuracy.

\subsection{Scalability}
We compared COVAP’s scalability with other GC schemes using 3 DNNs. The experiments were conducted on four clusters of 8, 16, 32 and  64 GPUs. Since we did not need to focus on training accuracy here, we used Cifar-10 as the dataset for ResNet-101 and VGG-19. The result is shown in Fig. 11, where the linear scaling bar represents the upper limit speedup of DP. 

Fig. 11(a) shows that COVAP obtained speedups of only 5\% less than linear scaling under four different clusters in ResNet-101 training, compared with 30\% less of PowerSGD. Besides, \verb|AllReduce|-based GC schemes showed no degradation as the cluster size increased since \verb|AllReduce| has better scalability than \verb|AllGather|. For example, \verb|AllGather|-based Random-k and EFsignSGD obtained only 3.2x acceleration while the GPU number increased by 8x. COVAP scaled even better than other \verb|AllReduce|-based schemes because it essentially reduced the number of communication operations in each iteration. Another reason is that COVAP adjusted its compression ratio adaptively according to the change of $CCR$ while other GC schemes kept their compression ratios (or other hyperparameters) unchanged. Specifically, COVAP outperformed other schemes by 1.15x-9.03x on 64 GPUs, compared with 1.04x-3.02x on 8 GPUs.

Fig. 11(b) and Fig. 11(c) show the scalabilities of all schemes in training VGG-19 and Bert. Note that we could not scale Top-k, Random-k, DGC, EFsignSGD, and Ok-topk beyond 16 GPUs in training VGG-19 since \verb|AllGather| requires more memory than \verb|AllReduce| when cluster scaled and caused running out of memory. The comparison of Fig. 11(a) and 11(b) disclosed that the 3x longer communication overhead of VGG-19 than ResNet-101 degraded its scalabilities of all schemes, while COVAP preserved the best stability in all cases along with cluster scaling.

\section{Limitations}
The compression ratio of COVAP correlates intensely to the $CCR$ of DNN training. As shown in Table \uppercase\expandafter{\romannumeral1}, the $CCR$ of DNNs at a common network of 30-100 Gbps bandwidth is generally not too high (i.e., \textless 5). However, in worse network cases such as federated learning or edge computing, the $CCR$ would be much higher, whereas the sparsification schemes using a high compression ratio work well. In such cases, an exorbitant compression ratio of COVAP may lead to an accuracy decline because of worse staleness \cite{b30}. One of our future works is to handle this issue.

\section{Related works}

\setlength{\parindent}{0cm}
$\mathbf{Gradient}$ $\mathbf{compression}$ $\mathbf{in}$ $\mathbf{poor}$ $\mathbf{network}$. Many GC schemes \cite{b16,b17,b22,b29} are proposed for poor network environments (e.g., \textless 10Gbps bandwidth) where the communication bottleneck is particularly severe. Therefore, those works often focus on enlarging their compression ratios to reduce communication overhead as much as possible. In addition, several GC schemes \cite{b6,b15,b27} were proposed to improve security and privacy in federated learning since private information can be leaked through shared gradients.

\begin{table}
\setlength{\abovecaptionskip}{0.cm}
\setlength{\belowcaptionskip}{-0.cm}
\setlength{\leftskip}{-2pt}
  \caption{Stages of selected layers or tensors}
  \resizebox{86.5mm}{7mm}{
  \begin{tabular}{cccc}
    \hline
    Technique&Forward Pass&Gradient computation&communication\\
    \hline
    LayerDrop&discarded&discarded&discarded\\
    Freeze training&reserved&discarded&discarded\\
    COVAP&reserved&reserved&discarded\\
  \hline
\end{tabular}}
\end{table}

{
\setlength{\parindent}{0cm}
$\mathbf{LayerDrop}$ $\&$ $\mathbf{Freeze}$ $\mathbf{training}$. Fan \cite{b9} proposed a form of structured dropout called LayerDrop, which selects and drops several layers of DNN in training. The motivation of LayerDrop is to regularize very deep Transformers while stabilizing its training. It also makes the network robust to subsequent pruning. Freeze Training is a helpful technique widely adopted in the object detection field. Since some parts of pre-training weights are universal, such as the backbone, we do not need to compute the gradients and update the weights of those layers. In PyTorch, it is easy to use Freeze Training by setting the $require\_grad$ of those layers to False. 
}

{
\setlength{\parindent}{0.4cm}
Similar to these two techniques, COVAP also discards some stages of DP (Note that these two techniques are not GC schemes). Table \uppercase\expandafter{\romannumeral8} shows their difference. Besides the difference in discarded stages, COVAP adopts a different discard granularity: \verb|Tensor|, instead of layer in LayerDrop and Freeze training.
}

\section{Conclusion}
{
\setlength{\parindent}{0.4cm}
COVAP is a novel gradient compression scheme combining Overlapping for distributed data parallel deep learning training. COVAP adaptively reduces the communication overhead approach to the computation time, making the communication almost fully hidden while introducing close-to-zero compression overhead. Empirical results for data parallel training of real-world deep learning models on Alibaba Cloud ECS instances show that COVAP achieves near-linear scaling in all experiments and significantly improves the training speed by up to 15.39x than existent gradient compression schemes while reaching similar model accuracy to the baseline.
}

\section*{Acknowledgment}
{
\setlength{\parindent}{0.4cm}
This work is supported in part by Science and Technology Innovation 2030 - Major Project (No. 2022ZD0119104). The opinions, findings and conclusions expressed in this paper are those of the authors and do not necessarily reflect the views of the funding agencies or the government.
}

\end{document}